\documentclass[aps,prl,amsfonts,twocolumn,showpacs,floatfix]{revtex4}

\usepackage{graphicx}
\usepackage{amsmath}
\usepackage{bm}

\begin{document}

\title{Temperature-dependent disorder and magnetic field driven disorder: experimental observations for  doped GaAs/AlGaAs quantum well structures}

\author{N.V.Agrinskaya}
\author{V.A. Berezovets}

\author{V.I.Kozub}

\affiliation{A. F. Ioffe Institute of Russian Academy of Sciences,
194021 Saint Petersburg, Russia}

\begin{abstract}

 We report experimental studies of conductance and magnetoconductance of GaAs/AlGaAs quantum well structures where both wells and barriers are doped by acceptor impurity Be. Temperature dependence of conductance demonstrate a non-monotonic behavior at temperatures around 100 K. At small temperatures (less than 10 K) we observed strong negative magnetoresistance at moderate magnetic field which crossed over to positive magnetoresistance
at very strong magnetic fields and was completely suppressed with an increase of temperature. We ascribe these unusual features to effects of temperature and magnetic field on a degree of disorder. The temperature dependent disorder is related to charge redistribution between different localized states with an increase of temperature. The magnetic field dependent disorder is also related by charge redistribution between different centers, however in this case an important role is played by the  doubly occupied states of the upper Hubbard band, their occupation being sensitive to magnetic field due to on-site spin correlations. The detailed theoretical model is present.

\end{abstract}

\maketitle

\section{Introduction}
It is a common wisdom to consider disorder of different sorts as
an important source of electric resistance for conductors on
both sides of metal-insulator transition. At the same time it is
typical to consider the degree of the disorder as some given
feature of a specific sample not depending on external
perturbations like temperature or magnetic field. This
statement seems to be indeed undisputable for standard metals
where the disorder has a structural character and is related to
structural defects, in particular, to some impurities. However it
is questionable for semiconductors where the main source of
disorder is related to charged defects. The charge states
of the defects can depend on temperature due to charge
redistribution between different carrier states. Then, one should
also take into account an existence of double-occupied carrier
states (so-called upper Hubbard band). Due to spin correlations
within such double-occupied states their occupation numbers can
also depend on external magnetic field and thus the contribution of these states to disorder becomes magnetic-field dependent. However, as far as we
know, the concept of temperature-dependent disorder was first
introduced only in 2000 in \cite{Maslov} in a course of an attempt
to explain features of the apparent metal-insulator transition in
2D structures (its nature is still far from complete
understanding). Then, the similar attempt lead to a concept of
magnetic-field driven disorder (that is, disorder dependent on external magnetic field) (\cite{Pudalov}, \cite{Agrin})
while a discussion of these two concepts was given in
\cite{Agrin}. At the same time both temperature dependent disorder
and magnetic field dependent disorder - while being discussed for Si MOSFETs -
still were not studied in detail for systems where the disorder is related to well
defined localized states. To our
opinion, the attractive system is related to a structure of
quantum wells where both wells and barriers are doped. The properly selected doping of both wells and barriers facilitates a creation of double occupied states in  a
controlled way due to tunneling of carriers from the barriers to
the wells. In our earlier work we reported
long-term relaxations within the response to external magnetic
field for GaAs/AlGaAs structures doped by Be which we explained as
the Coulomb glass effects \cite{Agrin2}, \cite{Agrin3}. An
important ingredient of our explanation was related to an effect
of magnetic field on the charge distribution within the Coulomb
glass. While the papers \cite{Agrin2}, \cite{Agrin3} described the effects of non-stationary relaxations related to pulses of external magnetic field, in this  paper we are going to present our results of
both experimental and theoretical studies of stationary temperature dependence of resistance and
magnetoresistance for GaAs/AlGaAs structures (similar to the ones
studied in \cite{Agrin2}, \cite{Agrin3}).
Note, however, that in these previous investigations we studied
dilute system of dopants within the barrier where the collective effects were neglected.
Thus the densities of impurity states were completely controlled by
pair configurations of charged states within the barrier and closest charged states within the well.
In our case we deal with significantly larger acceptors concentration than in \cite{Agrin3} which
leads to significant band broadening related to the collective effects, in other words - to the effects of random potential.

Our results
include, first, an unusual non-monotonic temperature dependence of resistance
exhibiting it saturation in course of temperature increase which
terminate initial resistance growth (typical for phonon mechanism
of resistance). The importance of this result is evidenced by a statement given in \cite{PRB}: "Understanding of non-monotonic behavior of temperature dependence of resistance $R(T)$ is a central issue in the correlated 2D carriers in clean semiconductors in studies of metallic state and metal-insulator transition".
Second, we observed an unusually strong negative
magnetoresistance at small temperatures which could not be
ascribed to standard interference mechanisms. We explain these unusual features as a direct manifestation of temperature-dependent
disorder and magnetic field dependent disorder. The
theoretical model for the structures under study is in a
qualitative agreement with the experimental data.

\section{Experiment}
We studied GaAs quantum wells structures (with a width 15 nm) separated by
Al$_{0.3}$Ga$_{0.7}$As barriers (with a width 25 nm). The growth procedure was described in detail
in~\cite{Agrin2}. Central regions (with a width of 5 nm) of both wells
and barriers were $p$-doped with Be (concentration $(3-7)10^{17}$
atoms/cm$^3$), which is close to the critical concentration for metal-insulator transition). Note that in the bulk material the metal-insulator transition takes place at impurity concentration close to
$(1-2)10^{18}$atoms/cm$^3$,~\cite{Agrin2}. The contacts were made by 2 min burning at 450$^\circ$
C in deposited gold containing 3\% of Zn. The samples were shaped as
Hall bars. The resistance was determined from the voltage between the
voltage probes at a fixed current of 1- 10 nA.
 The samples were relatively low-Ohmic ($10^5 -
10^6$~Ohms/$\square$ at 4~K). In our opinion, it is due to the fact
that the impurity band formed by $A^+$ centers is rather close to the
valence band. We checked that for all our measurements the $I-V$ curves were
linear within the temperature region 4.2-1.35 K.

The temperature dependencies of the resistance are shown on fig.1 for 3 samples with different concentrations of dopants. Separately, fig.2 presents the low temperature behavior of conductivity for sample 5-482. As it can be seen from Fig.1, at  some regions of high temperatures we observe a decrease of conductance with an increase of temperature. Although typically such a decrease would be ascribed to electron-phonon scattering, such a mechanism is not expected to be effective for samples impurities concentrations as high as we have in our samples since the impurity scattering strongly dominates the phonon scattering. When the temperature becomes lower the conductivity increases with temperature increase which is explained by activation of the holes from the upper Hubbard band to the valence band  (the activation energy $\varepsilon_2$ for different samples is $\sim (5-7)meV$). At higher temperatures several samples exhibited steep increase of conductivity with high activation energies $\varepsilon_1\sim (20-30)meV$. We believe that such a behavior is related to ionization of carriers from the lower Hubbard band  (see Fig. 1 ).

At very low temperatures (up to 1.2 K) the conductance is supported by weakly localized states within the upper Hubbard band. The weak localization of the carriers responsible for the conductivity is evidenced by a logarithmic temperature dependence of the conductivity (see Fig. 2).

\begin{figure}[htbp]
\centering
\includegraphics[width=8cm]{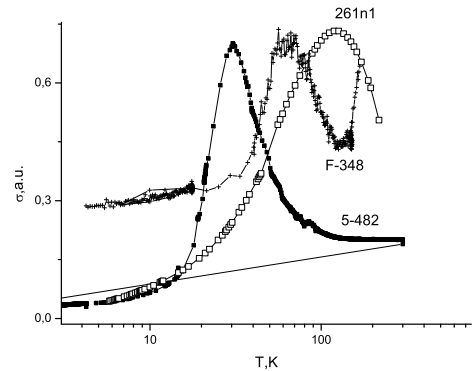}
\caption{Temperature dependencies of conductivities ( in arbitrary units) of three samples with different impurity (Be) concentrations in wells and barriers (f-348 -$(3)10^{17}$
/cm$^3$), f-261 -$(5)10^{17}$ /cm$^3$), 5-482 -$(7)10^{17}$ /cm$^3$)  }
\label{fig1}
\end{figure}

\begin{figure}[htbp]
\centering
\includegraphics[width=8cm]{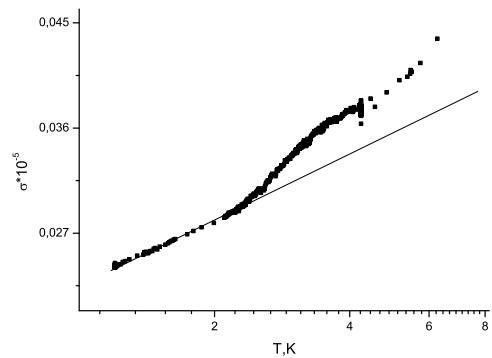}
\caption{Temperature dependencies of conductivity  for sample 5-482 -$(7)10^{17}$ /cm$^3$) at low temperatures  }
\label{fig2}
\end{figure}

 The magnetoresistance curves were measured at high magnetic fields (up to 15 T).
 The observed  behavior is strongly different for different temperatures (at temperature region 1.6 - 10 K we observe a crosssover from giant negative MR at 1,6Ê up to weakly positive MR at 10 Ê, see Fig.3.
\begin{figure}[htbp]
\centering
\includegraphics[width=8cm]{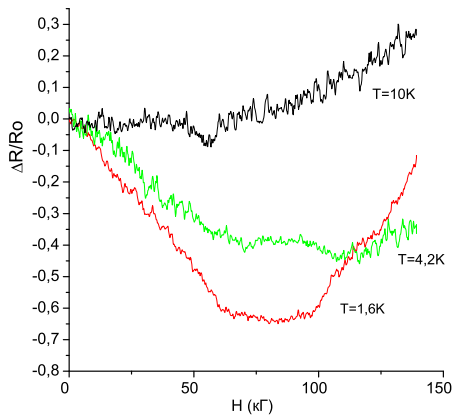}
\caption{Magnetoresistance curves for sample 5-482 -$(7)10^{17}$ /cm$^3$ for different temperatures. }
\label{fig3}
\end{figure}

\section{Discussion}

a) Temperature-dependent disorder.

Let us consider an effect of temperature dependent disorder. It is
related to a lifting of non-linear screening of localized states
by some carriers (see, e.g., \cite{Agrin}). The non-linear screening implies a capture of
the carrier by the (charged) localized state. Depopulation of the
corresponding trap due to activation of the carrier to the
delocalized state inevitably leads to a creation of charged center
giving a contribution to the (static) disorder. As for the trapped
carriers, in our case we deal with two types of centers. 1) $A^+$
center where a hole is trapped by neutral $A^0$ center. In this
case the activation of a hole leaves a neutral $A^0$ center. Thus in this case
the temperature increase leads to a {\it  partial
suppression} of the disorder rather than to increase of disorder.
2) ${\tilde A}^0$ centers created by an acceptor within the
barrier screened by a hole situated near the interface. Here
indeed an activation of a hole leaves (deeply localized) negative
$A^-$ center contributing to the total disorder potential. Note
that in the latter case depopulation of ${\tilde A}^0$ center
leads simultaneously to a creation of a free hole (increasing the
conductance) and of a charged scatterer (decreasing the
conductance). While an addition to the conductance is controlled by a
charge of a hole, the strength of the scatterer can be different
depending on a position of the scatterer and on collective effects.
Indeed, the charged acceptors in the barrier are not
strongly separated from each other thus creating a complex
potential relief.

Note that in our previous studies \cite{Agrin3} we exploited a model \cite{Larsen} developed for
dilute system of dopants within the barrier where the collective effects were neglected.
Thus the densities of states for $A^+$ and ${\tilde A}^0$ subbands were completely controlled by
pair configurations involving $A^-$ center within the barrier and closest $A^0$ center within the well.
In this case the density of states for $A^+$ centers was constant within some well-defined energy interval.
In our case we deal with significantly larger acceptors concentration than in \cite{Agrin3} which
leads to significant band broadening related to the collective effects, in other words - to the effects of
random potential.

In particular, one can expect that the strength
of the resulting scatterer depends on the energy level of
the hole trapped by the corresponding potential well. Namely, the deeper is the hole level (and, thus,
the smaller is the activation exponent), the stronger should be
the potential responsible for the trapping. Thus, the stronger
is the corresponding scatterer. One can expect that, at least at
the lowest approximation, the cross-section of scattering is
$\propto |V_s|^2$ where $V_s$ is a magnitude of the scattering
potential. In its turn, one also expects that the energy of the
localized state is $\varepsilon \propto |V_s|$. Thus one concludes
that the relaxation rate related to the corresponding scatterer is scaled
with the energy of the state bounded to the center as $\tau^{-1}
\propto |\varepsilon|^2$.

 Now let us consider redistribution of the holes supplied by the dopants within the barriers (
 the corresponding concentration within the well being $N_b$) over 3 subbands: 1) the valence band, 2) $A^+$ band, 3) $\tilde A^0$ band.
Naturally one has for the partial concentrations:
\begin{eqnarray}\label{subbands}
N_{A^+} = \int_{(A^+)} {\rm d} \varepsilon g_{A^+} (\varepsilon) F_0 (\varepsilon) \nonumber\\
N_{A^+} =
\int_{({\tilde A}^0)} {\rm d} \varepsilon g_{{\tilde A}^0} (\varepsilon) F_0 (\varepsilon) \nonumber\\
N_{v} =
\int_{\varepsilon_v} {\rm d} \varepsilon g_v (\varepsilon) F_0 (\varepsilon) \nonumber \\
N_{A^+} + N_{{\tilde A}^0} + N_v = N_b; \hskip1cm F_0 =
\frac{1}{\exp \frac{(\varepsilon - \Omega)}{T} + 1)}
\end{eqnarray}
Here for each of the subbands the integration is carried over the
corresponding energy band. Actually here we have the equation for
the chemical potential $\Omega$.

We will discriminate between 2 temperature regions. The first one
corresponds to low temperatures when the chemical potential nearly
coincides with the Fermi level $\varepsilon_F$. We will assume
that the Fermi level is near the bottom of $A^+$ band and,
correspondingly, near the top of the ${\tilde A}^0$ band. It can
be related to the fact that the magnitude of the binding potential
for ${\tilde A}^0$ states is in average larger than the binding
energy of $A^+$ center. In this temperature region one can neglect
a finite population of the valence band. Note that the temperature
behavior of resistance (see Fig. 1 ) at $T \rightarrow 0$
demonstrates nearly metallic behavior. Thus one can conclude that
the states at the Fermi level are delocalized, although the
mobility edge (we specify it as the energy separating strongly and weakly localized states) is rather close to $\varepsilon_F$ since the
conductance is rather small.

At high enough temperatures the carriers are activated to the
valence band. They are also activated to the upper states within
the $A^+$ band which are expected to be well delocalized (being
far away from the mobility edge). For a simplicity we will not
discriminate between these well delocalized states of $A^+$ band
and the valence band (assuming that the bands are well
overlapped). Correspondingly, we will still define some boundary
energy separating "well delocalized states" and "poor delocalized
states" ascribing it to the bottom of the "effective valence
band".

Let us consider now redistribution of the holes within this
simplified picture including only ${\tilde A}^0$ band and the
(modified) valence band. (see Fig.4).

\begin{figure}[htbp]
\centering
\includegraphics[width=9cm]{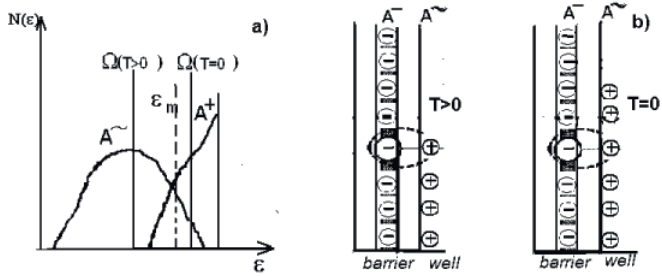}
\caption{a) Energy spectrum including upper Hubbard band (emerging to the valence band) and ${\tilde A}^0$ band.  $\varepsilon_m$ specify a position of the mobility edge. The positions of the chemical potential at $T = 0$ and finite $T$ (to describe  experimental behavior it is of the order of 100 K). b) Energy spectrum of ${\tilde A}^0 $ states and of $A^-$ states (corresponding to ionized ${\tilde A}^0$ states) for two
values of temperature mentioned above. To emphasize the complex dipolar character of ${\tilde A}^0$ states some of them are confined by elliptical curves.}
\label{fig4}
\end{figure}

 While at $T = 0$ chemical potential
coincides with the Fermi level, at finite temperature it is
shifted downwards:
\begin{equation}
\delta \Omega = T\frac{g_v}{g_{{\tilde A}^0}} \exp (-\varepsilon_F/T)
\end{equation}
where $g_v$ is density of state for valence band. Thus we have for
the partial depopulation of ${\tilde A}^0$ band:
\begin{equation}
\Delta N_{{\tilde A}^0} = \frac{g_{{\tilde A}^0}g_v}{g_{{\tilde
A}^0}}T \exp (- \varepsilon_F/T)
\end{equation}
Now let us consider a behavior of conductance controlled by the
valence band:
\begin{eqnarray}\label{sigmaA0}
\sigma = \frac{e^2\delta N_v}{  \frac{m}{{\bar \tau}_{{\tilde A}^0}(\varepsilon = \varepsilon_F)}N_{A^+}(T = 0) + B}\nonumber\\
B = \int^{\mu}_{\varepsilon_F} {\rm d}\varepsilon  \frac{m}{{\bar
\tau}_{{\tilde A}^0}(\varepsilon)} g_{{\tilde A}^0}
\end{eqnarray}
Here $1/{\bar \tau}(\varepsilon)$ is a contribution of single
scatterer to the relaxation rate.  As it is seen, initially (at small
$T$) $B \simeq 0$ and the value of $\sigma$ steeply increases with
temperature increase due to exponential increase of $N_v$. The
situation becomes different when the chemical potential is deep
within ${\tilde A}^0$ band. In this case it is seen that
the integration over $\varepsilon$ in equation for $B$ is
controlled by the upper limit since $1/{\bar \tau} \propto
\varepsilon^2$. In its turn, the position of chemical potential is
exponentially shifted which leads to steep increase of $B$ and
finally it dominates the first term in the denominator
(originating from residual scatterers existing at $T=0$) . In
general, since $g_{{\tilde A}^0}(\Omega - \varepsilon_F) = \delta
N_v$, the integral can be estimated as $e^2 N_v(m/{\bar \tau}(\Omega)$.
Thus, after initial exponential increase of $\sigma$, it is
followed by a steep decrease due to steep increase of $1/{\bar
\tau}$. Then, an increase of $1/{\bar \tau}$ with temperature
increase is naturally restricted by some effective value $(1/{\bar
\tau})_{max}$ which leads to a saturation of $\sigma(T)$ at the
value
\begin{equation}
\sigma \simeq \frac{e^2}{m (1/{\bar \tau}_{max})}
\end{equation}

b) Magnetic field driven disorder.

The main effect of the external magnetic field $H$ on the relation between $A^+$ and ${\tilde A}^0$ centers
is related to the Zeeman energy $\mu_B g H$  (where $\mu_B$ is Bohr magneton while $g$ is $g$-factor) which should be paid for a creation of $A^+$ center
due to on-site spin correlation on the center   (see Fig.5).

\begin{figure}[htbp]
\centering
\includegraphics[width=7cm]{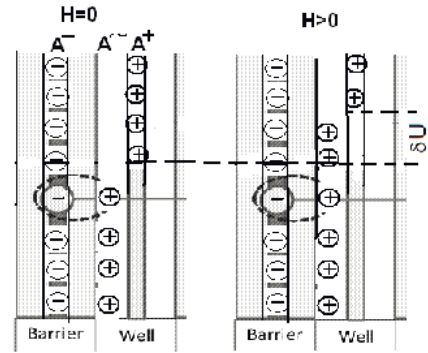}
\caption{Relation between the number of $A^+$ states and ${\tilde A}^0$ states for different values of external magnetic field}
\label{fig5}
\end{figure}

 We will denote the corresponding energy as $\delta U_{A^+}$; actually it is an addition to the Hubbard energy. Note that $\delta U_{A^+}$ is literally equal to the Zeeman energy only at low temperatures
$T << \mu_B g H$. Otherwise we have $\delta U_{A^+} \simeq (\mu_BgH)^2/4 T$.
The effect is expected to be pronounced for small temperatures. Thus, to estimate the effect, we can
take the densities of states at the Fermi level. One expects that
the magnetic field decreases the concentration of $A^+$ centers as
\begin{equation}\label{dis1}
\frac{\delta N_{A^+}}{N_A} = - \delta U_{A^+} g_{A^+}
\end{equation}
If we would deal with a diluted system, the effect would arise from a difference between scattering efficiencies for the dipole potential
created by $A^+$ and $A^-$ centers and the dipole potential corresponding to ${\tilde A}^0$ center. Since the latter has significantly smaller intercharge distance, one could expect that the ratio given by Eq.\ref{dis1} describes a relative
suppression of the disorder by magnetic field. In particular, it could describe a relative increase of metallic-like
conductance observed at low temperature. However one can expect the effect to be much stronger. Indeed, as it was noted above,
the low-temperature conductance is undoubtedly close to the metal-insulator transition. Thus due to a shift of the mobility edge a decrease of disorder  can lead to much stronger effect on the conductance than predicted by Eq.\ref{dis1}. In particular, one can even expect a magnetic-field driven metal-insulator transition.

\section{Conclusions}

  A study of temperature dependent disorder and magnetic field driven disorder for the case of model structures where the disorder originated from well defined localized states is presented. As an example of such systems we used GaAs/AlGaAS quantum wells where both wells and barriers were doped by acceptor impurity Be. Such a doping ensures a partial occupation of the upper Hubbard band where the doubly occupied states are sensitive to external magnetic field due to on-site spin correlations. The main observed features included unusual steep decrease of conductance with temperature increase which can not be ascribed to standard phonon scattering and strong negative magnetoresistance at moderate fields which was suppressed by temperature increase. We developed a consistent theoretical model which is in semiquantitative agreement with experimental data.

\section{Acknowledgements}

This work was partly supported by Russian Foundation, Grant N 13-02-00169. We are also indebted to Yu.M.Galperin for reading the manuscript and many valuable remarks.

\end{document}